\documentclass[a4paper]{article}

\usepackage{amsmath,amssymb,graphicx}
\usepackage{subfigure}

\usepackage{lscape}
\usepackage{mathrsfs}

\usepackage{algorithmic}
\usepackage[ruled]{algorithm}
\usepackage{graphicx}\usepackage{dcolumn}\usepackage{bm}\usepackage{amsmath}\usepackage{amscd}\usepackage{amsfonts}\usepackage{extarrows}
\usepackage{multirow}
\usepackage{verbatim}
\usepackage[colorlinks,
            linkcolor=red,
            anchorcolor=red,
            citecolor=red,
            ]{hyperref}
\usepackage{algorithmic}
\usepackage[ruled]{algorithm}
\makeatletter
\newenvironment{breakablealgorithm}
  {
   \begin{center}
     \refstepcounter{algorithm}
     \hrule height.8pt depth0pt \kern2pt
     \renewcommand{\caption}[2][\relax]{
       {\raggedright\textbf{\ALG@name~\thealgorithm} ##2\par}%
       \ifx\relax##1\relax 
         \addcontentsline{loa}{algorithm}{\protect\numberline{\thealgorithm}##2}%
       \else 
         \addcontentsline{loa}{algorithm}{\protect\numberline{\thealgorithm}##1}%
       \fi
       \kern2pt\hrule\kern2pt
     }
  }{
     \kern2pt\hrule\relax
   \end{center}
  }
\makeatother

\def\ds{\displaystyle}
\usepackage{color}

\newtheorem{lemma}{Lemma}
\newtheorem{theorem}{Theorem}
\newtheorem{corollary}{Corollary}

\newtheorem{proposition}{Proposition}

\newtheorem{remark}{Remark}

\def\a{{\bf a}}
\def\b{{\bf b}}
\def\hb{{\bf b}}
\def\c{{\bf c}}

\def\e{{\bf e}}

\def\hr{{\bf r}}

\def\x{{\bf x}}
\def\y{{\bf y}}

\def\0{{\bf 0}}
\def\1{{\bf 1}}
\def\2{{\bf 2}}
\def\3{{\bf 3}}
\def\4{{\bf 4}}
\def\5{{\bf 5}}
\def\6{{\bf 6}}
\def\7{{\bf 7}}
\def\8{{\bf 8}}
\def\9{{\bf 9}}

\def\bn{\begin{definition}}
\def\en{\end{definition}}
\def\ba{\begin{array}}
\def\ea{\end{array}}
\def\be{\begin{equation}}
\def\ee{\end{equation}}
\def\bd{\begin{description}}
\def\ed{\end{description}}
\def\bu{\begin{enumerate}}
\def\eu{\end{enumerate}}
\def\bi{\begin{itemize}}
\def\ei{\end{itemize}}

\def\bt{\begin{theorem}}
\def\et{\end{theorem}}
\def\bp{\begin{proposition}}
\def\ep{\end{proposition}}
\def\bc{\begin{corollary}}
\def\ec{\end{corollary}}
\def\bo{\begin{proof}}
\def\eo{\end{proof}}
\def\bx{\begin{example}}
\def\ex{\end{example}}
\def\br{\begin{remark}}
\def\er{\end{remark}}
\def\bl{\begin{lemma}}
\def\el{\end{lemma}}

\def\ds{\displaystyle}
\newcommand{\qed}{$\hfill{\blacksquare}$}

\usepackage{longtable}

\title{ Randomized Row and Column Iterative Methods with a Quantum Computer  }

\author{Changpeng Shao\thanks{Academy of Mathematics and Systems Science, Chinese Academy of Sciences, Beijing 100190, P. R. China. (cpshao@amss.ac.cn). }
~and
Hua Xiang\thanks{School of Mathematics and Statistics, Wuhan University, Wuhan 430072, P. R. China.  Corresponding author (hxiang@whu.edu.cn).  }
}

\begin{document}
\date{}
\maketitle

\begin{abstract}

We consider the quantum implementations of the two classical iterative solvers for a system of linear equations, including the Kaczmarz method which uses a row of coefficient matrix in each iteration step, and the coordinate descent method
which utilizes a column instead. These two methods are widely applied in big
data science due to their very simple iteration schemes. In this paper we use the block-encoding technique and propose fast quantum implementations for these two approaches, under the
assumption that the quantum states of each row or each column can be efficiently prepared. The
quantum algorithms achieve exponential speed up at the problem size over the classical versions,
meanwhile their complexity is nearly linear at the number of steps. 

\end{abstract}
%



\section{Introduction}

The classical solvers for a linear system of equations $A\x=\b$ are generally categorized into two types:   direct methods and   iterative methods. The latter is usually more practical in the realm of large-scale system of equations.
Among the iterative methods, the Kaczmarz method and the coordinate descent method are popular due to the simplicity and efficiency.
The Kaczmarz method was first discovered 
in 1937 \cite{Kaczmarz}, and was rediscovered in the field of image reconstruction by
Gordon, Bender and Herman in 1970 \cite{Gordon-Bender-Herman-ART} under the appellation
algebraic reconstruction technique (ART).
The Kaczmarz method uses a row of coefficient matrix in each iteration, while the coordinate descent method
utilizes a column instead.
These two methods seek to solve different problems: 
the coordinate descent method converges to a least squares solution generally; 
the Kaczmarz method calculates a minimum-norm solution for a consistent system of equations and
exhibits cyclic convergence for an inconsistent problem \cite{ElfvingHansenNikazad_NA}.
They can be generalized to many variants (see \cite{XiangZhang} and the references therein).
The advantage of these two methods lies in the fact that at each iteration they only need access to an individual row (or column) rather than the entire coefficient matrix.
Due to the simplicity, they have numerous applications in the fields ranging from computer tomography to image processing and digital signal processing, especially the big data science.

We review the iteration schemes in the following.
Assume that $A$ is an $n$-by-$n$ matrix. For any $1\leq i\leq n$, denote the $i$-th row of $A$ as $\a_i^T$, and the $i$-th component of $\b$ as $b_i$.
Let $\x_0$ be an arbitrary initial approximation to the solution of $A\x=\b$.
For $k\geq 0$, randomly choose an $i_k\in\{1,\ldots,n\}$,
the Kaczmarz iteration updates the solution $\x_k$ as follows:
\be \label{Kaczmarz-Iteration-eq1}
\x_{k+1} = \x_k - \left( \frac{\a_{i_k}^T\x_k}{\|\a_{i_k}\|} - \frac{b_{i_k}}{\|\a_{i_k}\|} \right) \frac{\a_{i_k}}{\|\a_{i_k}\|} ,
\ee
%
%
which is equivalent to the Gauss-Seidel method on $AA^T \y = \b$, where $A^T \y=\x$.
Geometrically, $\x_{k+1}$ is the orthogonal projection of $\x_k$ onto the hyperplane $\a_{i_k}^T\x = b_{i_k}$.
In each iteration step, only one row of the coefficient matrix is needed. It is also called the row-action method. Another names, such as component-solution method, cyclic projection or successive projection, are used on certain occasions.

Correspondingly, we have the column-action method. Let $\c_j$ be the $j$-th column of $A$, and $\e_j$ the $j$-th column of the unit matrix.
The column-action method reads
\begin{equation}\label{eqn:CD}
 \x_{k+1} = \x_k + \frac{ \c_{j_k}^T( \b-A \x_k )}{\| \c_{j_k} \|^2} \e_{j_k} ,
\end{equation}
where $j_k$ is a random number from $\{1,\ldots, n\}$.
It is equivalent to the randomized Gauss-Seidel method on $A^T A\x= A^T \b$.
In each iteration step the approximate solution is obtained by a minimization in one coordinate direction, then it is also called the
coordinate descent method. Define $ {\bf r}_k = \b - A \x_k$. 
This iteration method can be reexpressed as
\be\ba{lll} \vspace{.2cm} \label{eqn:CD-new}
\x_{k+1}  &=&  \ds \x_k + \frac{ \c_{j_k}^T  {\bf r}_k } {\| \c_{j_k} \|^2} \e_{j_k}, \\
\hr_{k+1} &=&  \ds \hr_k -  \frac{ \c_{j_k} \c_{j_k}^T  } {\| \c_{j_k} \|^2} \hr_k.
\ea\ee

Suppose that the size of the problem is $n$ and the number of required iterations to solve the problem is $T$,
then the complexity of classical iteration algorithm is usually polynomial at $n$ and linear at $T$.
Due to the quantum no-clone theorem, an iteration algorithm  is usually not easy to implement in a quantum computer.

For instance, in \cite{Rebentrost-Newton}, Rebentrost et al. proposed a quantum gradient and Newton's method to solve polynomial optimization.
Compared to the classical gradient or Newton's method, this quantum algorithm achieves exponential speedup at $n$.
However, the complexity exponentially depends on $T$. The main reason is as follows, taking Newton's method as an example.
One critical step of classical Newton's method is to solve a linear system.  The coefficient matrix
(i.e., the Hessian matrix) $H$ of the linear system depends on $\x_k$,  the computed result in step $k$.
In a quantum computer, we only have the quantum state of $\x_k$, which is a unknown quantum state.
To solve the linear system by HHL algorithm, one critical step is the Hamiltonian simulation of $H$.
In \cite{Rebentrost-Newton}, a similar idea to quantum principal component analysis \cite{Lloyd-PCA} is used to compute the
Hamiltonian simulation of $H$ by viewing it as a unknown density matrix. Together with quantum phase estimation,
the linear system can be solved efficiently.
However, the Hamiltonian simulation of a unknown density operator requires $O(t^2/\epsilon)$ copies
of this density operator, where $t$ is the evolution time and $\epsilon$ is the precision.
Since $H$ depends on $|\x_k\rangle$,
we need to prepare at least $O(t^2/\epsilon)$ copies of $|\x_k\rangle$. In other words, with at least $O(t^2/\epsilon)$ copies of $|\x_k\rangle$,
we can prepare $|\x_{k+1}\rangle$. As a result, we need exponentially copies of the initial state to prepare $|\x_T\rangle$.

In \cite{Kerenidis-IPM}, Kerenidis and Prakash considered a quantum version of interior-point method to solve semi-definite programming and linear programming.
One critical step of classical interior-point method is also to solve a linear system.
Different from the idea used above, they output the classical information of $|\x_k\rangle$ at each step of iteration. By doing so
$|\x_k\rangle$ becomes a known quantum state. To read out the classical information of a quantum state by quantum tomography
requires at least $O(n)$ steps. Therefore, the complexity of their quantum algorithm is polynomial at $n$.
The dependence on $T$ is polynomial due to the error caused in quantum tomography in each step.

These are two typical examples of quantum iteration algorithms. They cannot outperform classical iteration algorithms both at $n$ and $T$.
The quantum iteration algorithm is a simulation of classical iteration algorithm in a quantum computer,
so it seems especially hard to achieve a speedup at $T$. However, we still expect the quantum iteration algorithm
to achieve high speedup at $n$ and has a reasonable dependence on $T$ meanwhile.
Therefore, an ideal simulation of iteration algorithm in a quantum computer should have the
complexity polynomial-logarithm at $n$ and linear or polynomial at $T$.
In this paper, we will give such implementations to the Kaczmarz method and the coordinate descent method.
The main idea is to use block-encoding technique \cite{Chakraborty-block-encoded} to implement the procedures (\ref{Kaczmarz-Iteration-eq1}), (\ref{eqn:CD-new})
by unitary operators. These unitary operators are explicitly constructed and efficiently implemented in the quantum computer. Moreover, they can help us
overcome the calculations of the inner product of quantum states. As a result, we can show that
the Kaczmarz method and the coordinate descent method can be implemented in a quantum computer in time $O(T(\log n))$.

Throughout this paper, we use the following notations.

\vspace{-3mm}

\begin{enumerate}
\item[i.] A quantum state of the form $\sum_{i=0}^{2^k-1}\alpha_i |i\rangle |\psi_i\rangle $ will be simply denoted by
$\alpha_0|0\rangle^{\otimes k}|\psi_0\rangle + |0^\bot\rangle^{\otimes k}|\cdots \rangle$, when we only concern about $|\psi_0\rangle$ and neglect the garbage state.

\vspace{-2mm}

\item[ii.] We denote ${\rm SWAP}_{i,j}$ as the swap operator that swaps the $i$-th qubit and the $j$-th qubit.
\end{enumerate}

\section{The quantum Kaczmarz algorithm}

Assume that the quantum state of $\a_t$ can be prepared efficiently in the quantum computer, such as by qRAM \cite{Giovannetti}.
So there is an efficiently implemented unitary operator $V_t$ such that $V_t|0\rangle = |\a_t\rangle$.
Based on the iteration formula (\ref{Kaczmarz-Iteration-eq1}), without loss of generality, we suppose that $\|\a_t\|=1$ for all $t$.
By equation (\ref{Kaczmarz-Iteration-eq1}), we have
\be \label{updating rule-quantum}
|\x_{k+1}\rangle  \varpropto  \|\x_k\| \Big( I_n - |\a_{i_k}\rangle \langle \a_{i_k}| \Big) |\x_k\rangle + b_{i_k} |\a_{i_k}\rangle.
\ee

For any row index $t$, define a unitary operator
\be\label{unitary-matrix}
U_t = \left[
        \begin{array}{cc} \vspace{.2cm}
          I_n -  |\a_t\rangle \langle \a_t| &~~ |\a_t\rangle \langle \a_t| \\
           |\a_t\rangle \langle \a_t|     &~~ I_n - |\a_t\rangle \langle \a_t| \\
        \end{array}
      \right]=I_2\otimes (I_n -  |\a_t\rangle \langle \a_t|) + X\otimes |\a_t\rangle \langle \a_t|,
\ee
where $X$ is the Pauli-X matrix. That is,
\be
U_t = (I_2\otimes V_t) (I_2\otimes (I_n -  |0\rangle \langle 0|) + X\otimes |0\rangle \langle 0|) (I_2\otimes V_t^\dag).
\ee
By qRAM assumption, $V_t$ is efficiently implemented in the quantum computer, so is $U_t$.

By equation (\ref{unitary-matrix}), $U_t$ can be viewed as a control operator: if the second
register is $|\a_t\rangle$, then apply $X$ to the first register; if the second
register lies in the orthogonal complement space of $|\a_t\rangle$, then do nothing to the first register.
The general architecture of these kind of unitaries was studied in \cite{Gilyen-Quantum-singular-value-transformation}.

The basic idea of the quantum implementation of Kaczmarz iteration is as follows:
Suppose that we have the following state that contains the quantum information of $|\x_k\rangle$
\be
|X \rangle = \sqrt{p} \, |0\rangle |\x_k \rangle + \sqrt{1-p} \, |1\rangle  |\cdots \rangle.
\ee
Let $\beta^2 + \gamma^2 =1$, then we can prepare
\be
|\psi \rangle = {\rm SWAP}_{1,2} \Big( \beta |0\rangle|X\rangle + \gamma  |1\rangle |0\rangle  |\a_t\rangle \Big)
 = |0\rangle  \Big( \beta \sqrt{p} \, |0\rangle |\x_k\rangle + \gamma |1\rangle |\a_t\rangle \Big)
 + \beta\sqrt{1-p}\, |1\rangle  |0\rangle |\cdots \rangle.
\ee
As to the first term, direct calculation yields that
\be
 U_t \Big( \beta \sqrt{p} \,|0\rangle |\x_k\rangle + \gamma |1\rangle |\a_t\rangle \Big) = |0\rangle \otimes
\Big( \beta\sqrt{p} (I_n - |\a_t\rangle \langle \a_t|) |\x_k\rangle + \gamma |\a_t\rangle \Big)
+ \beta\sqrt{p}\langle \a_t | \x_k \rangle |1\rangle |\a_t\rangle .
\ee
Properly choosing the parameters $\beta,\gamma$, for example, $\beta=\|\x_k\|\delta, \gamma=b_t \sqrt{p} \delta$ for some $\delta$ to
ensure $\beta^2 + \gamma^2 =1$,
then the first term of $|\psi\rangle$ is a state proportional to the right hand side of equation \eqref{updating rule-quantum}.

The explicit procedure to implement Kaczmarz iteration is stated as follows.

\begin{breakablealgorithm}
\caption{\bf The quantum Kaczmarz method }
\label{Simulate the Kaczmarz method in a quantum computer 2}
\begin{algorithmic}[1]
\STATE Randomly choose a unit vector $\x_0$ such that its quantum state can be prepared in time $O(\log n)$.
Set $k=0$ and $\mu_k = 1$. The state can be expressed in the following general form
      \begin{equation} \label{extend-state-solu}
       |X_k\rangle = \frac{\|\x_k\|}{\mu_k} \, |0\rangle^{\otimes k} \otimes |\x_k\rangle + |0^\bot\rangle^{\otimes k}|\cdots \rangle.
      \end{equation}
\STATE Randomly choose a $t_k\in\{1,\ldots,n\}$.
       Define $ \beta_{t_k}^2 = \ds \frac{\mu_k^2}{\mu_k^2 + b_{t_k}^2}$,
       $\gamma_{t_k}^2=1-\beta_{t_k}^2$ and $\mu_{k+1} = \ds \frac{\mu_k}{\beta_{t_k}}$.
\STATE Apply $(I_2^{\otimes k}\otimes U_{t_k}) {\rm SWAP}_{1,k+1}$ to
       $\beta_{t_k} |0\rangle|X_k\rangle + \gamma_{t_k}|1\rangle |0\rangle^{\otimes k}|\a_{t_k}\rangle$,
       then we obtain
       \[
       |X_{k+1}\rangle = \frac{\|\x_{k+1}\|}{\mu_{k+1}} \, |0\rangle^{\otimes (k+1)} \otimes |\x_{k+1}\rangle +  |0^\bot\rangle^{\otimes (k+1)}|\cdots \rangle.
       \]
\STATE Set $k=k+1$, and go to step 2 until converges.
\end{algorithmic}
\end{breakablealgorithm}

In step 3, we calculate that
\be\ba{lll} \vspace{.2cm}
&& \ds I_2^{\otimes k}\otimes U_{t_k} \Big( \frac{\beta_{t_k}  \|\x_k\|}{\mu_k}  |0\rangle^{\otimes k} |0\rangle  |\x_k\rangle
+ \gamma_{t_k} |0\rangle^{\otimes k} |1\rangle |\a_{t_k}\rangle + |0^\bot\rangle^{\otimes k} |0\rangle |\cdots \rangle  \Big) \\\vspace{.2cm}
&=& \ds |0\rangle^{\otimes (k+1)} \otimes \Big( \frac{\beta_{t_k}\|\x_k\|}{\mu_k} \Big(I_n -  |\a_{t_k}\rangle \langle \a_{t_k}| \Big)|\x_k\rangle
 + \gamma_{t_k} |\a_{t_k}\rangle \Big)  + |0^\bot\rangle^{\otimes (k+1)} |\cdots \rangle \\
&=& \ds  \frac{\beta_{t_k} }{\mu_k} |0\rangle^{\otimes (k+1)} \otimes \Big(  \|\x_k\| \Big(I_n -  |\a_{t_k}\rangle \langle \a_{t_k}| \Big)|\x_k\rangle
 + b_{t_k} |\a_{t_k}\rangle \Big)  + |0^\bot\rangle^{\otimes (k+1)}  |\cdots \rangle,
\ea\ee
where we use the fact that  $\gamma_{t_k} = \sqrt{ 1-\beta_{t_k}^2 } = \beta_{t_k} b_{t_k} /\mu_k$ in the last step.

Similar to the classical Kaczmarz method,
algorithm \ref{Simulate the Kaczmarz method in a quantum computer 2}
is also simple to implement in a quantum computer.
Let $\x_k$ be the result obtained by the classical Kaczmarz method in the $k$-th step.
Then the first term of $|X_k\rangle$ defined in equation (\ref{extend-state-solu})
contains all the information of $\x_k$ precisely, i.e., $\|\x_k\||\x_k\rangle$.

\bt \label{thm2}
Assume that $|\a_t\rangle$ is prepared in $O(\log n)$ for any $t$.
In algorithm \ref{Simulate the Kaczmarz method in a quantum computer 2},
for any $k\geq 1$, we have
\be
\mu_k^2 = 1 + \sum_{i=1}^{k-1} b_{t_i}^2.
\ee
The complexity to prepare $|X_k\rangle$ is $O(k\log n)$.
\et
{\em Proof.} By definition in step 2 of algorithm \ref{Simulate the Kaczmarz method in a quantum computer 2},
\[
\mu_{k+1}^2 = \frac{\mu_k^2}{\beta_{t_k}^2} = \mu_k^2 +b_{t_k}^2.
\]
Since $\mu_0=1$, we have
\[
\mu_k^2 = 1 +  \sum_{i=1}^{k-1}  b_{t_i}^2.
\]

Assume that the complexity to prepare $|X_k\rangle$ is $ \tau_k $, then
the complexity for $|X_{k+1}\rangle$ in step 3 is $ \tau_k + O(\log n)$, since the  time for preparing $|a_{t_k}\rangle$ is $O(\log n)$.
Thus, $ \tau_{k+1} = \tau_k + O(\log n)$. Since $\tau_0=O(\log n)$, we have
$\tau_k = O(k\log n)$.
\qed

\vspace{.2cm}

Strohmer et al. \cite{StrohmerVershynin_JFAA09} sample a row in a random fashion with probability proportional to the
2-norm of that row at each iteration, and prove an exponential expected convergence rate of randomized Kaczmarz method.
For simplicity, we assume  a uniform sampling in step 2.
Then the expectation reads
$$
\mathbb{E} [\mu_k^2] = 1 + (k-1) \mathbb{E} [b_{t_i}^2] = 1+ \frac{k-1}{n} \|\b\|^2_2 .
$$
If $\|\b\|_\infty=O(1)$, then $\mathbb{E} [\mu_k] =O(\sqrt{k})$.

The classical information of the solution $\x_{k}$ is stored in the first term of  $|X_{k}\rangle$.
For some problems in machine learning, such as   data classification, the final output extracts certain global information, rather than each component of a state.
For example, to estimate the inner product between $\x_{k}$ and the vector $\c$, we can
first prepare the quantum state $|\c\rangle$ of $\c$, then apply the swap test to estimate
the inner product between $|X_{k}\rangle$ and $|0\rangle^{\otimes k}|\c\rangle$. This returns an $\epsilon'$-approximate of
$\|\x_{k}\| \langle \x_{k}|\c\rangle/\mu_k$ in time $O(k(\log n)/\epsilon')$.
Thus, by setting $\epsilon' = \epsilon /\mu_k\|\c\|$, we will obtain an $\epsilon$-approximate of $\x_{k}\cdot\c$
in time $O(k(\log n)\mu_k\|\c\|/\epsilon)$.

\section{The quantum coordinate descent algorithm}

With a bit abuse of notations, in the following we use $\a_j$ to denote the $j$-th column of $A$.
Assume that the quantum state of $\a_j$ can be efficiently prepared. That is, there exist unitary operators $S_j$
such that $S_j^\dag |j\rangle = |\a_j\rangle$ for any $j$.

The coordinate descent method can be implemented as follows:
(1) Randomly choose an initial guess $\x_0$ and set $\hr_0 = \hb - A \x_0$.
(2) Randomly choose a $t_k\in\{1,\ldots,n\}$ and update
\be\ba{rll} \vspace{.2cm} \label{update rule:CD}
\x_{k+1}   &=& \ds \x_{k} + \frac{\a_{t_k}^T \hr_k}{\|\a_{t_k}\|^2} \e_{t_k}, \\
\hr_{k+1} &=& \ds \Big(I_n - \frac{\a_{t_k} \a_{t_k}^T}{\|\a_{t_k}\|^2} \Big) \hr_k.
\ea\ee
For convenience, we assume that $\|\a_t\|=1$ for any $t$.
Using quantum states, we can rewrite (\ref{update rule:CD}) as
\be\ba{rll} \vspace{.2cm} \label{update rule:CD-new}
|\x_{k+1}\rangle  &\varpropto& \ds \|\x_k\| \, |\x_k\rangle + \|\hr_k\| \, |{t_k}\rangle\langle \a_{t_k}|\hr_k\rangle , \\
|\hr_{k+1}\rangle &\varpropto& \ds \|\hr_k\| \, (I_n - |\a_{t_k}\rangle \langle \a_{t_k}| ) |\hr_k\rangle.
\ea\ee

Before implementing the procedure (\ref{update rule:CD-new}) in the quantum computer, we state some ideas below.
Firstly, we consider the update of the residual.
The basic idea is the same as algorithm \ref{Simulate the Kaczmarz method in a quantum computer 2}.
Suppose that the residual of the $k$-th step is encoded in the state
\be
|R_k\rangle = \|\hr_k\|\, |0\rangle^{\otimes k} |\hr_k\rangle + |0^\bot\rangle^{\otimes k}|\cdots \rangle.
\ee
Apply $(I_2^{\otimes k} \otimes U_{t_k}) {\rm SWAP}_{1,k+1}$ to $|0\rangle |R_k\rangle$, then
\be\ba{lll} \vspace{.2cm}
(I_2^{\otimes k} \otimes U_{t_k}) {\rm SWAP}_{1,k+1} |0\rangle |R_k\rangle
&=& |0\rangle^{\otimes (k+1)} \|\hr_k\| (I_n-|\a_{t_k}\rangle \langle \a_{t_k}|) |\hr_k\rangle + |0^\bot\rangle^{\otimes (k+1)} |\cdots \rangle \\ \vspace{.2cm}
&=& \|\hr_{k+1}\| \, |0\rangle^{\otimes (k+1)}   |\hr_{k+1}\rangle + |0^\bot\rangle^{\otimes (k+1)} |\cdots \rangle \\
&=& |R_{k+1} \rangle.
\ea\ee
This is in fact the algorithm \ref{Simulate the Kaczmarz method in a quantum computer 2} with initial vector $\hr_0$ and  $\mu_k=1$ for all $k$.

Secondly, we consider the update of the approximate solution.
Since $\langle t|S_t = \langle \a_t|$, we have
\be
|\x_{k+1}\rangle \varpropto \|\x_k\| \, |\x_k\rangle + \|\hr_k\| \, |{t_k}\rangle\langle {t_k}| S_{t_k}|\hr_k\rangle.
\ee
Algorithm \ref{Simulate the Kaczmarz method in a quantum computer 2} is not applicable to the above procedure directly.
Some modifications are required. The following are some basic ideas.

Suppose that  the approximate solution $\x_k$ and the corresponding residual $\hr_k$ are encoded in the following states, respectively
\be
|\widetilde{X}_k \rangle = \frac{\|\x_k\|}{\mu} |0\rangle |\x_k\rangle + |0^\bot\rangle|\cdots\rangle,
\ee
and
\be
|\widetilde{R}_k \rangle = \| \hr_k\| \, |0\rangle  S_{t_k} | \hr_k\rangle  + |0^\bot\rangle|\cdots\rangle.
\ee
We then show how to combine them to generate a state that contains $|\x_{k+1} \rangle$.

Introduce two auxilla qubits, and prepare
\be
|\phi_1\rangle = \beta |00\rangle |\widetilde{X}_k \rangle + \gamma |10\rangle |\widetilde{R}_k \rangle,
\ee
where $\beta^2+\gamma^2 = 1$.
For any $t$, define
\be
W_t = \left[
        \begin{array}{ccc} \vspace{.2cm}
          I_n & 0 & 0 \\ \vspace{.2cm}
          0 & I_n - |t\rangle\langle t| & |t\rangle\langle t| \\
          0 & |t\rangle\langle t|     & I_n - |t\rangle\langle t| \\
        \end{array}
      \right].
\ee
Apply ${\rm SWAP}_{2,3}W_{t_k}$ to $|\phi_1 \rangle$ to prepare
\be\ba{lll}  \vspace{.2cm}
|\phi_2\rangle &=& {\rm SWAP}_{2,3} W_{t_k} |\phi_1\rangle \\  \vspace{.2cm}
&=& {\rm SWAP}_{2,3} (\beta |00\rangle |\widetilde{X}_k \rangle + \gamma |01\rangle |t_k\rangle \langle t_k  |\widetilde{R}_k \rangle  + |10\rangle|\cdots\rangle) \\
&=& \ds |00\rangle \Big(\frac{\beta\|\x_k\|}{\mu} |0\rangle  |\x_k\rangle + \gamma  \|\hr_k\| \, |1\rangle |t_k\rangle \langle t_k| S_{t_k} |\hr_k\rangle\Big)
+ |0^\bot\rangle^{\otimes 2} |\cdots\rangle.
\ea\ee

Define $G_k =  \left[
       \begin{array}{rr} \vspace{.2cm}
         c & s \\
         -s  & c \\
       \end{array}
     \right]$, where $c^2+s^2 =1$. Apply $I_4 \otimes G_k \otimes I_n $ to $|\phi_2\rangle$, then we obtain
\be
|\phi_3\rangle = (I_4 \otimes G_k \otimes I_n) |\phi_2\rangle
= |000\rangle \Big(  \frac{c\beta\|\x_k\|}{\mu} |\x_k\rangle + s \gamma  \|\hr_k\| \, |t_k\rangle \langle t_k| S_{t_k} |\hr_k\rangle \Big)
+ |0^\bot\rangle^{\otimes 3} |\cdots\rangle.
\ee
We can properly choose the parameters $c,s,\beta,\gamma$, such that the first term of $|\phi_3\rangle$ is proportional to
\be
\|\x_k\| \, |\x_k\rangle +  \|\hr_k\| \, |t_k\rangle \langle t_k| S_{t_k} |\hr_k\rangle = \|\x_{k+1} \| \, |\x_{k+1}\rangle .
\ee

With the preparations above, we can present the quantum coordinate descend algorithm as follows,
where
\be
G_k= \frac{1}{\sqrt{k+2}} \left[
       \begin{array}{cc} \vspace{.2cm}
         \sqrt{ k+1 } & 1 \\
         - 1  & \sqrt{ k+1 } \\
       \end{array}
     \right].
\ee

\begin{breakablealgorithm}
\caption{\bf The quantum coordinate descend method }
\label{The coordinate descend method in a quantum computer}
\begin{algorithmic}[1]
\STATE Randomly choose a unit vector $\x_0$ such that its quantum state can be prepared in time $O(\log n)$.
Assume that $\hr_0=\b-A\x_0$ has unit norm and its quantum state is prepared in time $O(\log n)$.
Set $k=0$. Denote
      \be\ba{lll}\label{CD-alg-step1} \vspace{.2cm}
       |X_k\rangle &=& \ds \frac{\|\x_k\|}{k+1} \, |0\rangle^{\otimes 2k} \otimes |\x_k\rangle + |0^\bot\rangle^{\otimes 2k}|\cdots \rangle, \\
       |R_k\rangle &=& \|\hr_k\|\, |0\rangle^{\otimes k} |\hr_k\rangle + |0^\bot\rangle^{\otimes k} |\cdots \rangle.
      \ea\ee
\STATE Randomly choose a $t_k\in\{1,\ldots,n\}$.
\STATE Apply $(I_2^{\otimes (2k+1)}\otimes G_k\otimes I_n)(I_2^{\otimes 2k}\otimes W_{t_k}) {\rm SWAP}_{2,2k+2} {\rm SWAP}_{1,2k+1}$ to
\be  \label{ancilla-state}
\sqrt{\frac{k+1}{k+2}} \, |0 0 \rangle|X_k\rangle
+  \sqrt{\frac{1}{k+2}} \, |1 0 \rangle (I^{\otimes 2k} \otimes S_{t_k}) |0\rangle^{\otimes k} |R_k\rangle ,
\ee
then we obtain $|X_{k+1}\rangle$.
\STATE Apply $(I_2^{\otimes k} \otimes U_{t_k}) {\rm SWAP}_{1,k+1}$ to $|0\rangle |R_k\rangle$ to generate $|R_{k+1}\rangle$.
\STATE Set $k=k+1$, and go to step 2 until converges.
\end{algorithmic}
\end{breakablealgorithm}

We explain the update of approximate solution in details.
The state (\ref{ancilla-state}) in step 3 is denoted as $|\psi_0\rangle$, that is,
\be\ba{lll}  \vspace{.2cm}
|\psi_0\rangle 
&=& \ds
\sqrt{\frac{k+1}{k+2}} \, |00\rangle \Big(\frac{\|\x_k\|}{k+1} \, |0\rangle^{\otimes 2k} |\x_k\rangle
                        + |0^\bot\rangle^{\otimes 2k}|\cdots \rangle\Big) \\
&& \ds +\,  \sqrt{\frac{1}{k+2}} \, |10\rangle
\Big(\|\hr_k\|\, |0\rangle^{\otimes 2k} S_{t_k}|\hr_k\rangle + |0^\bot\rangle^{\otimes 2k}|\cdots \rangle\Big).
\ea\ee
Apply ${\rm SWAP}_{2,2k+2} {\rm SWAP}_{1,2k+1}$ to $|\psi_0\rangle$, we obtain
\be\ba{lll}  \vspace{.2cm}
|\psi_1\rangle &=& \ds
\sqrt{\frac{k+1}{k+2}} \,  \Big(\frac{\|\x_k\|}{k+1} \, |0\rangle^{\otimes 2k}  |00\rangle|\x_k\rangle
                        + |0^\bot\rangle^{\otimes 2k}|00\rangle|\cdots \rangle\Big) \\ \vspace{.2cm}
&& \ds +\,  \sqrt{\frac{1}{k+2}} \,
\Big(\|\hr_k\|\, |0\rangle^{\otimes 2k} |10\rangle S_{t_k}|\hr_k\rangle + |0^\bot\rangle^{\otimes 2k}|10\rangle|\cdots \rangle\Big) \\
&=& \ds
|0\rangle^{\otimes 2k} \otimes \Bigg( \sqrt{\frac{k+1}{k+2}}  \frac{\|\x_k\|}{k+1} \,  |00\rangle|\x_k\rangle
+ \sqrt{\frac{1}{k+2}} \|\hr_k\| \, |10\rangle  S_{t_k} |\hr_k\rangle \Bigg) + |0^\bot\rangle^{\otimes 2k} |\cdots \rangle.
\ea\ee
Apply $I_2^{\otimes 2k}\otimes W_{t_k}$ to $|\psi_1\rangle$ to get
\be
|\psi_2\rangle =
|0\rangle^{\otimes (2k+1)} \otimes \Bigg( \sqrt{\frac{k+1}{k+2}}  \frac{\|\x_k\|}{k+1} \,  |0\rangle|\x_k\rangle
+ \frac{\|\hr_k\|}{\sqrt{k+2}}  |1\rangle |t_k\rangle\langle t_k|S_{t_k} |\hr_k\rangle \Bigg) + |0^\bot\rangle^{\otimes (2k+1)} |\cdots \rangle.
\ee
Apply $I_2^{\otimes (2k+1)}\otimes G_k\otimes I_n$ to $|\psi_2\rangle$, then we get
\be\ba{lll} \vspace{.2cm}
|\psi_3\rangle &=& \ds
|0\rangle^{\otimes 2(k+1)} \otimes \Bigg( \frac{k+1}{k+2}  \frac{\|\x_k\|}{k+1} \,  |\x_k\rangle
+ \frac{\|\hr_k\|}{k+2} |t_k\rangle\langle t_k|S_{t_k} |\hr_k\rangle \Bigg) + |0^\bot\rangle^{\otimes 2(k+1)} |\cdots \rangle \\ \vspace{.2cm}
&=& \ds \frac{\|\x_{k+1}\|}{k+2} \, |0\rangle^{\otimes 2(k+1)} \otimes |\x_{k+1}\rangle +  |0^\bot\rangle^{\otimes 2(k+1)}  |\cdots \rangle \\
&=& |X_{k+1}\rangle.
\ea\ee

Similar to the proof of theorem \ref{thm2}, we obtain the following result

\bt \label{thm3}
In algorithm \ref{The coordinate descend method in a quantum computer}
the complexity to prepare $|X_k\rangle$ is $O(k\log n)$.
\et

\begin{remark}{\rm
Since $\|\x_0\|=1$, now assume that $\|\x_k\|\leq k+1$ and $\|\hr_0\|=1$, then by equation (\ref{update rule:CD-new}),
\[
\|\x_{k+1}\|^2 = \|\x_k\|^2 + 2 \|\hr_k\| \|\x_k\| \langle \a_{t_k}|\hr_k\rangle \langle t_k|\x_k\rangle
+\|\hr_k\|^2\langle \a_{t_k}|\hr_k\rangle^2
\leq (k+1)^2 + 2(k+1)+1 = (k+2)^2.
\]
Therefore, by induction the equation (\ref{CD-alg-step1}) is well-defined.
}\end{remark}

\begin{remark}{\rm
By definition, $\hr_0=\b-A\x_0$, with a suitable choice of $\x_0$ we can make sure that it has unit norm.
Even if it does not has unit norm, algorithm \ref{The coordinate descend method in a quantum computer}
still works. Let $\rho$ be a parameter such that $\rho\|\hr_0\|\leq 1$.
In  algorithm \ref{The coordinate descend method in a quantum computer}, it suffices to change $|R_0\rangle$ into
$|\widehat{R}_0\rangle = \rho \|\hr_0\|\, |0\rangle |\hr_0\rangle + |0^\bot\rangle  |\cdots \rangle$.
By equation (\ref{update rule:CD-new}),
$\|\hr_{k+1}\| \, |\hr_{k+1}\rangle = \|\hr_k\| \, (I_n - |\a_{t_k}\rangle \langle \a_{t_k}| ) |\hr_k\rangle.$
Therefore, $|R_k\rangle$ used in algorithm \ref{The coordinate descend method in a quantum computer} is simply changed into
$|\widehat{R}_k\rangle = \rho \|\hr_k\|\, |0\rangle^{\otimes (k+1)} |\hr_k\rangle + |0^\bot\rangle^{\otimes (k+1)} |\cdots \rangle.$
Here the reason to use $k+1$ ancilla qubits is the extra one ancilla qubit introduced in $|\widehat{R}_0\rangle$.
Since $\rho\|\hr_0\|\leq 1$, we have $\rho \|\hr_k\| \leq 1$ for all $k$.
}\end{remark}

\begin{remark}{\rm
For the update of residual we can use the linear combinations of unitaries (LCU) \cite{Childs-LCU}.
It contains many applications in quantum computing, such as quantum simulation \cite{Berry-HS,Childs-LCU},
quantum linear solver \cite{Childs-linear-solver}.
Let $U_0,\ldots,U_{m-1}$ be $m$ unitary operators, and $\alpha_0,\ldots,\alpha_{m-1}$ be $m$ positive
real numbers.  Set $s=\sum_j \alpha_j$.
Assume that $V$ is a unitary operator that maps $|0\rangle^{\otimes \log m}$ to $\frac{1}{\sqrt{s}} \sum_j \sqrt{\alpha_j} |j\rangle$.
Given  a quantum state $|\psi\rangle$, the technique of LCU can compute $\sum_j \alpha_j U_j |\psi\rangle$.
In our case, if we choose $U_0=I_n$, $U_1= I_n - 2 |\a_{t_k}\rangle\langle \a_{t_k}|$ and $\alpha_0=\alpha_1=1/2$,
then we can prepare $(I_n - |\a_{t_k}\rangle\langle \a_{t_k}|)|\psi\rangle$.
}\end{remark}

\section{Conclusions}

The quantum implementation of a general iterative method is usually very challenging. The typical
quantum iteration methods in \cite{Kerenidis-IPM,Rebentrost-Newton} cannot outperform classical iteration methods in both the problem size $n$ and the iteration number $T$.
We therefore switch to two special iterative methods: the Kaczmarz method and the coordinate descent method.
For solving linear systems, these two methods may not be absolutely superior to other iterative algorithms.
But due to their simple structures that only one row (or column) is accessed at each iteration, they are
very popular in certain areas, such as the big data science.  In this paper, we show that the quantum versions of Kaczmarz method and coordinate descent method
also have simple implementation structures in a quantum computer. Moreover, the efficiency is exponentially better in the problem size than their classical counterparts.



Our quantum iterative linear solvers are different from other quantum linear solvers
\cite{Childs-linear-solver,Harrow-Hassidim-Lloyd-HHL,Wossnig-Zhao-Prakash}.
The methods in this paper are independent of Hamiltonian simulation and quantum phase estimation, and also shed some lights on the quantum implementation of iterative methods.
The only assumption on our algorithms is the requirement of a qRAM, by which we can efficiently extract the row or column of the coefficient matrix.
 One drawback of our quantum iterative linear solvers is the number of ancilla qubits.
Since the iteration is generally not a unitary procedure, we need to introduce ancilla qubits to change it
into a unitary one. It remains an open problem to find better ways to reduce the number of ancilla qubits.

\end{document}